\documentclass[8pt,twocolumn,a4paper]{article}
  \usepackage[dvips]{graphicx}
 \usepackage{amssymb}
 \usepackage[square]{natbib}
 \usepackage[font=small,labelfont=bf]{caption}

\begin{document}

%
%

\title{Resonant and non-resonant whistlers-particle interaction in the radiation belts}
%

%

%
%



\author{Enrico Camporeale\\
  Center for Mathematics and Computer Science (CWI)\\ 
  1098 XG, Amsterdam, Netherlands\\
  \texttt{e.camporeale@cwi.nl; }}
 \date{\today}

 \twocolumn[
  \begin{@twocolumnfalse}
    \maketitle
    \begin{abstract}
     We study the wave-particle interactions between lower band chorus whistlers and an anisotropic tenuous population of
relativistic electrons. 
We present the first direct comparison of first-principle Particle-in-Cell (PIC) simulations with a quasi-linear diffusion code, in this context.
In the PIC approach, the waves are self-consistently generated by a temperature anisotropy instability that quickly
saturates and relaxes the system towards marginal stability.
We show that the quasi-linear diffusion and PIC results have significant quantitative mismatch 
in regions of energy/pitch angle where the resonance condition is not satisfied.
Moreover, for pitch angles close to the loss cone the diffusion code overestimates the scattering, particularly
at low energies.
This suggest that higher order nonlinear theories should be taken in consideration 
in order to capture non-resonant interactions, resonance broadening, and to account for scattering at angles close to $90^\circ$.\\
    \end{abstract}
  \end{@twocolumnfalse}
]

%
%



%
%

%

\section*{Introduction}
Resonant wave-particle interactions play a fundamental role in 
space plasma physics.
In the radiation belts, energetic electrons that are potentially harmful to satellites are 
subject to pitch angle scattering
and local acceleration due to cyclotron and Landau resonances \citep{thorne10,summers_book}. 
 The non-conservation of the 
adiabatic invariants of motion leads to de-trapping and to scattering into the loss-cone,
where particles eventually precipitate into the ionosphere.
One of the major challenge, in a space weather perspective, is the accurate prediction of the timescale associated 
with the loss mechanisms of such energetic particles.\\
The standard theoretical framework for the modeling of wave-particle interactions
is represented by quasi-linear theory, initially developed in the seminal papers
by \citet{kennel66}, and broadly used in the context of cosmic ray acceleration \citep{jokipii66, kulsrud69,schlickeiser89}, 
tokamaks \citep{hazeltine81}, and
radiation belt physics \citep{lyons72,summers98}.
The quasi-linear procedure describes wave-particle interactions by means of a 
diffusion equation in pitch angle and energy for the particle distribution function,
by expanding particle orbits around their unperturbed trajectory in the Vlasov-Maxwell equations \citep{swanson_book}.
The complexity of wave-particle interactions is thus dramatically reduced to a diffusive process, and all the 
physical information is lumped into the diffusion coefficients, usually defined as a function of particles' pitch angle and energy.
Following the quasi-linear paradigm, the derivation and numerical calculation of diffusion coefficients for several kinds of plasma waves 
has been the focus of a great part of radiation belt physics 
in recent years \citep{albert04,summers05,glauert05,shprits06,mourenas12}.
Multidimensional diffusion codes have proved quite successful in studying the time evolution of the electron distribution  
function before, during and after a storm \citep{thorne13, tu13, miyoshi06, jordanova10, fok08, tao09, varotsou08, albert09b, shprits09, subbotin09,tu09, su11}.
Moreover, the quasi-linear diffusion coefficients have been recently tested against test-particle simulations 
for whistler-wave chorus and electromagnetic ion-cyclotron waves (EMIC),  
in \citet{tao11} and \citet{liu10}, respectively, founding an excellent agreement. 
\citet{tao12} have also reported the breakdown of the quasi-linear theory
predictions when, as expected, the wave amplitude is sufficiently large.\\
It is important to remind that the resonant quasi-linear theory employed in radiation belt studies is based on the following three approximations: the waves have random phase and small amplitude, and the particles
are in (either cyclotron or Landau) resonance with the wave spectrum \citep{lemons12}. 
Although not strictly required, most quasi-linear calculations have been carried over  by assuming a spectrum of waves derived by
the cold plasma linear theory, i.e. neglecting thermal effects. 
Wave damping/growth is generally neglected, since it increases the complexity of the derivation 
of the diffusion coefficients.
Finally, an accurate calculation of the diffusion coefficient
requires the detailed information on the wave power spectrum, which is generally
assumed as a Gaussian centered around a dominant mode \citep{horne05}.

In this paper, we present Particle-in-Cell (PIC) simulations and we focus on the wave-particle interactions 
between energetic electrons and whistlers generated by an anisotropic suprathermal relativistic population.
With such approach the wave spectrum is self-consistently generated and no further assumptions are required.
The resonant interactions between particles and a wave field that is growing in time due to an ongoing kinetic instability
has not been studied before in a self-consistent way, in terms of energy and pitch angle scattering.
We present, for the first time, a quantitative comparisons between PIC and Fokker-Planck simulations.
In this way we can directly assess the range of validity of the resonant quasi-linear approach, and its drawbacks.
It is worth noting that in the cosmic ray acceleration context, several authors have highlighted 
the weaknesses of standard quasi-linear diffusion, leading to the development of second-order quasi-linear theory
and weakly nonlinear theory (see, e.g., \citep{shalchi05,qin09}), where the particle orbit calculation takes in account the electromagnetic 
perturbation.
In particular, the standard quasi-linear theory fails to predict the correct scattering for large pitch angles ('the $90^\circ$ problem')
\citep{tautz08}. More recently, \citet{ragot12} has questioned the relative importance between 
resonant and non-resonant interactions in a turbulent magnetized plasma.
Finally, it is important to emphasize that one of the most relevant quantities for space weather predictions is the
particle lifetime. A standard estimate is based on the  inverse of the bounce-averaged diffusion coefficient 
evaluated at the equatorial loss cone angle, for different energies \citep{shprits06}.
Such estimate has been validated in \citet{albert09}, and parametrized in \citet{shprits07}.
The electron loss timescale varies from few hours to few days depending on the latitude distribution of wave power, 
the energy and the cold plasma parameter $\alpha^* = \Omega_{e}^2/\omega_{p}^2$ (with $\omega_{p}$ and $\Omega_{e}$ 
the electron plasma and equatorial cyclotron frequency, respectively).
However, we will show that quasi-linear diffusion tends to overestimate diffusion rates at small pitch angles:
this important result suggests to reconsider the standard estimates of particle lifetime.\\
We focus on the physics of whistler waves, which are routinely observed in the magnetosphere,
and are believed to play a dominant role for relativistic electron acceleration and precipitation \citep{horne03}.
Whistler waves can be generated by man-made VLF transmissions \citep{dungey63}, or as the result of anisotropic plasma injection
during a magnetic storm \citep{jordanova10}.
Indeed, equatorial whistler-mode chorus can be excited by cyclotron resonance 
with anisotropic 10-100 KeV electrons injected from the plasmasphere 
\citep{summers07}.
A statistical analysis of chorus excitation observed by THEMIS has recently been presented by \citet{li10}.
They have reported dayside lower-band chorus generated by anisotropic 10-100 KeV electrons. 
The  superposed epoch analysis performed at GEO orbit by \citet{macdonald08}, during 
geomagnetic storms, suggests that whistler wave growth is related to relativistic electron
enhancements, but they have not found instances where the whistler marginal stability condition is actually reached,
thus suggesting that the anisotropic suprathermal population is seldom strongly unstable, but rather in a condition
of marginal stability. Indeed, the long standing
scenario envisioned for radiation belt electrons implies a delicate equilibrium between losses due to pitch angle scattering into the
loss-cone, and enhanced wave activity due to kinetic instabilities triggered by anisotropic loss-cone distributions \citep{lyons73}.

\section*{Methodology}
We present one-dimensional Particle-in-Cell (PIC) simulations performed with the implicit code
Parsek2D \citep{markidis09,markidis10}. 
In order to simulate a situation relevant to the lower-band chorus generation in the radiation belt, we have 
chosen the following parameters. The background homogeneous magnetic field is $B_0 = 4\cdot10^{-7} T$, 
corresponding
to the equatorial value at $L\sim 4.3$, and it is aligned with the box (the assumption of homogeneous field is justified
because the timescale of the simulation is much shorter than the bounce period).
The cold plasma parameter is $\alpha^* = 0.104$.
The electron population has a density of 15 cm$^{-3}$, and it is composed for $98.5\%$
by a cold isotropic Maxwellian (1 eV), and for $1.5\%$ by an anisotropic relativistic bi-Maxwellian 
distribution $f(v_{||},v_\perp) \sim \exp\left[-\alpha_\perp \gamma-(\alpha_{||} - \alpha_\perp)\gamma_{||}\right]$ 
(with $\gamma = (1-v^2/c^2)^{-1/2}$, $\gamma_{||} = (1-v_{||}^2/c^2)^{-1/2}$, and
parallel and perpendicular refer to the background magnetic field) \citep{naito13,davidson89}.
We choose $\alpha_{||}=25$, and $\alpha_\perp=4$. The velocity distribution function
has standard deviations $\sqrt{\langle v_{||}^2 \rangle} = 0.175$, and $\sqrt{\langle v_\perp^2 \rangle} = 0.325$ (normalized 
to speed of light)
corresponding to nominal temperatures of 8 KeV and 30 KeV, respectively.
Thus, the initial anisotropy of the suprathermal component is $T_\perp/T_{||}= 3.75$.
We note that in order to accurately recover the quasi-linear pitch angle diffusion,
one has to ensure that the wavevector separation $\Delta k=2\pi/L$ is small enough,
such that each particle is subject to a relatively broad spectrum of modes.
In our simulations the box length $L=400 c/\Omega_e$, and the most dominant modes have wavelength
of about 1/200 of the box length. The number of grid points is 8,000.
The diffusion code employed in this paper is described in \citet{camporeale13}.
We solve the non-bounce-averaged Fokker-Planck equation in energy and pitch angle, with diffusion coefficient
evaluated as in \citet{summers05}. Mixed energy/pitch angle diffusion is included,
and standard boundary conditions are used (see \citet{camporeale13}).

\section*{Results}
\begin{figure}[h]
 \noindent\includegraphics[width=8 cm]{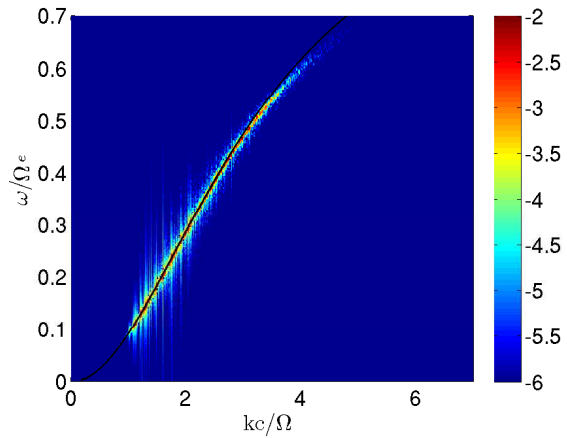}%
 \caption{Spectrogram of magnetic fluctuations, in logarithmic scale. The black line shows the cold plasma dispersion relation.
 Most of the wavepower is confined to $\omega\lesssim 0.5 \Omega_e$.\label{fig:spectrogram}}
 \end{figure}

In Figure \ref{fig:spectrogram}, we show the spectrogram of the magnetic fluctuations from PIC simulations, in logarithmic scale.
By virtue of the one-dimensional setup, the fluctuations are perpendicular to the background
field. The black line shows the cold plasma dispersion relation for equal parameters
(but, of course, without the suprathermal component), which is in very good agreement.

\begin{figure}[h]
 \noindent\includegraphics[width=8 cm]{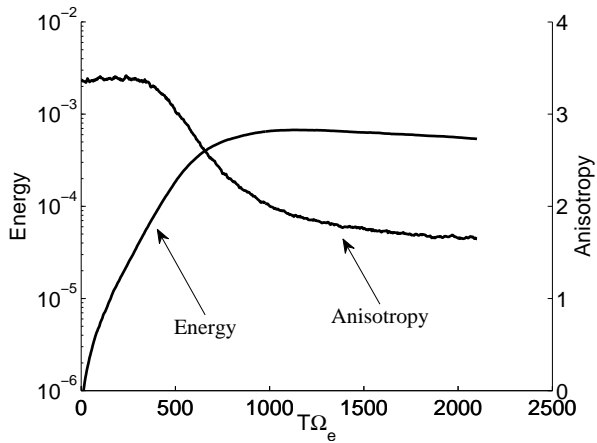}%
 \caption{Time development of energy (left axes, logarithmic scale), and temperature anisotropy (right axes, linear scale).
 Time is normalized to electron gyrofrequency.\label{fig:anisotropy}}
 \end{figure}

Note that most of the wave power is confined to $\omega/\Omega_e\lesssim 0.5$.
It is well known that temperature anisotropy instabilities have a 'self-destructing' character, in the sense
that the generated electromagnetic fluctuations reduce the anisotropy that drives the instability, and therefore a marginal
stability condition is usually rapidly reached \citep{camporeale08,gary14},
This effect is shown in Figure \ref{fig:anisotropy}. Electromagnetic energy (left axes) and anisotropy (right axes) are plotted as a function of time
(in electron gyroperiod units).
The reduction of anisotropy is indeed very strongly correlated to the linear growth phase of the instability,
roughly for $T\Omega_e<1000$.

The linear instability saturates at a large amplitude $\delta B/B_0 \sim 10\%$, and thus in the post-saturation regime
the quasi-linear theory might not be applicable. Since diffusion coefficients are very sensitive
to the magnetic wave power spectrum (but actually not much to the exact shape of the spectrum), a comparison between 
PIC simulations and a diffusion code raises the question of which magnetic spectrum to choose, since this is evolving in time during
the linear growth phase. We show results for two values of fluctuation amplitude: $\delta B/B_0 = 1\% $ and $2\%$ , which are
reached approximately at times $T\Omega_e = 200$ and 350, respectively, that is at the early stage of the linear growth.
We have tested that such values give the best agreement with PIC results, and as such our conclusions must be interpreted
as a 'best case scenario', in the case in which one employs the standard way of calculating the diffusion coefficients,
i.e. by neglecting wave growth, and using a chosen fluctuation amplitude. 
It is important to point out that larger or smaller values of $\delta B/B_0$ result in a larger disagreement.
Moreover, the PIC results suggest that it is reasonable to use a Gaussian spectrum centered
at $\omega/\Omega_e=0.2$ , with semi-bandwidth equal to 0.25.
\subsection*{Resonance curves}
Before commenting on the comparison between PIC and diffusion code results, it is useful to briefly revise few basic concepts
about wave-particle resonance and resonance curves.
A particle is in resonance with a wave with frequency $\omega$ and wavevector $k$ if the following relation is satisfied:
\begin{equation}\label{resonance}
 \omega- k v \cos\alpha = n\Omega_e\sqrt{1-v^2},
\end{equation}
which simply means that the relativistic gyrofrequency of the particle matches the Doppler-shifted wave frequency.
$n=0$ and $n=1,2,\ldots$ are respectively for Landau and cyclotron resonances.
Eq. (\ref{resonance}) produces to so-called resonance curves, that are ellipses in ($v_{||}, v_\perp$) space
on which the resonance condition is satisfied.
If the wave is confined within a certain range of frequencies (and wavevectors), as it is the case here (see Figure \ref{fig:spectrogram}),
Eq. (\ref{resonance}) can be used to calculate the minimum energy that is required for a particle with a given pitch angle
in order to fulfill the resonance condition. To this purpose, Eq. (\ref{resonance}) can be rewritten, for $n=1$, as
\begin{equation}\label{resonance2}
 \cos\alpha = \frac{\omega_M(E+1)-1}{k\sqrt{E^2+2E}},
\end{equation}
where $E$ is here the relativistic energy normalized to the rest mass, and $\omega_M$ the upper bound of the frequency range normalized
to $\Omega_e$.
For completeness, we recall that the wavevector $k$ can be calculated by using the cold plasma dispersion relation
for whistler waves:
\begin{equation}
 kc/\omega = \sqrt{1-\frac{\omega_p^2/\Omega_e^2}{\omega(\omega-\Omega_e)}}
\end{equation}

\subsection*{Comparison between PIC and diffusion code}
Figures 3, 4, and 5 show a direct comparison between PIC and diffusion codes, 
in terms of probability density function (pdf) of the suprathermal
species. Of course, such quantity is readily available in PIC simulations, and the statistics is here performed on 160,000 particles.
The pdf is not normalized, but rescaled such that its maximum value at initial time is equal to one.
Figure 3 shows the pdf as a function of pitch angle for energies $E=20$, 50, 100, 200 KeV. The black dashed line denotes
the initial condition. Red and blue lines show the results of the diffusion code for $\delta B/B_0=1\%$ and $2\%$, respectively, 
while PIC results are shown with black circles.
All the results are for time $T\Omega_e=1000$.

 \begin{figure}[h]
   \hspace{-1 cm}\noindent\includegraphics[width=9 cm, height = 8cm]{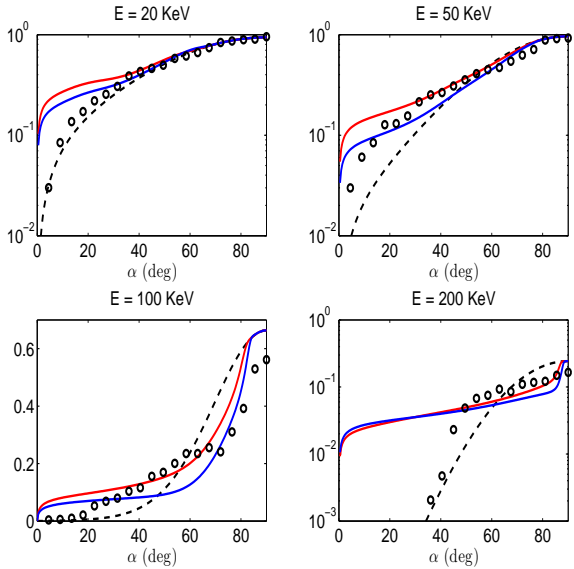}%
 \small\caption{ Comparison between PIC and diffusion code results. Probability density
 functions for energies E= 20, 50, 100, 200 KeV, as function of pitch angle $\alpha$ (in degrees).
 Black dashed lines are the initial condition. Red and blue lines show the results from diffusion code, with $\delta B/B_0=1\%$ and
 $2\%$, respectively. Black circles are for PIC results.
 All results are for $T\Omega_e = 1000$. Note that the bottom-left panel is shown with linear vertical scale, 
 while the others have logarithmic scale.\label{fig:diff_angle}}
 \end{figure}
 
The overall agreement is good, particularly for $\delta B/B_0=1\%$, but three features are evident. 
First, the diffusion code overestimates the diffusion at small pitch angles.
Second, the diffusion code does not capture scattering close to $90^\circ$ pitch angle. Note that the bottom-left panel
has a linear vertical axes to highlight such effect, while the other panels are presented in logarithmic scale.
Third, for each energy, there is a range in pitch angles where the diffusion code does not predict any diffusion, i.e.
solid and dashed lines overlap. This follows from the argument that we have presented above:
particles need to have a sufficient energy in order to fulfill the resonance condition.
This feature is even more evident in Figure 4.

 \begin{figure}[h]
  \hspace{-.3 cm}\noindent\includegraphics[width=9 cm, height = 8 cm]{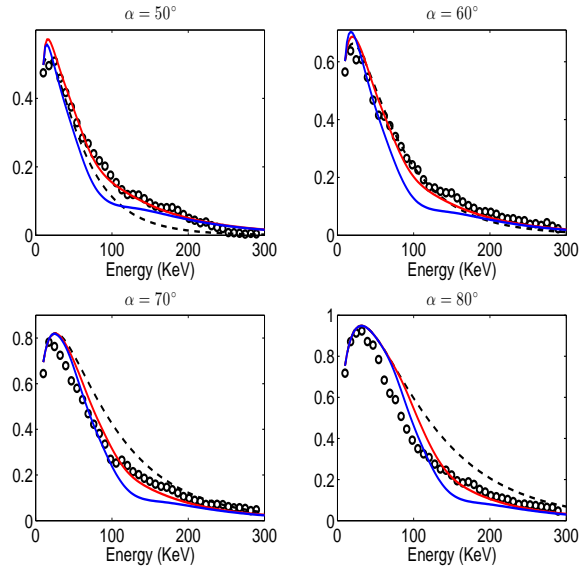}%
 \caption{Comparison between PIC and diffusion code results. Probability distribution
 functions for pitch angles $\alpha=50^\circ$, $60^\circ$, $70^\circ$, $80^\circ$ 
 as function of energy, at time $T\Omega_e = 1000$. Same legend as in Figure 3.\label{fig:diff_energy}}
 \end{figure}
 
Here we show (with same legend as in Figure 3) the pdf evolution as function of energy for pitch angles 
$\alpha=50^\circ$, $60^\circ$, $70^\circ$, $80^\circ$.
Again, the agreement between the codes is remarkably good, particularly for large energies, where the red lines and black circles
overlap almost exactly (remember, however, that we have cherry-picked an appropriate value for $\delta B/B_0$). And again, at smaller energies,
the diffusion code predicts little diffusion, while the PIC code results show a general decrease of the pdf with respect
to the initial value. This is very clear, for instance, in the bottom-right panel of Figure 4.
Finally, in Figure 5 we present the two-dimensional color plots of the particle distribution functions for PIC (top-left), 
and diffusion code
(top-right, $\delta B/B_0=1\%$), in logarithmic scale, at time $T\Omega_e=1000$.
Figure 5 reinforces the widespread viewpoint that, despite its numerous assumptions and limitations, 
the quasi-linear diffusion approach is indeed very powerful in capturing the overall features of resonant energy/pitch angle scattering.
On the other hand, the bottom panel of Figure 5 evidently emphasizes the downsides of the diffusion code. 
 \begin{figure}[h]
 \hspace{-0.5 cm}\noindent\includegraphics[width=9 cm, height = 9 cm]{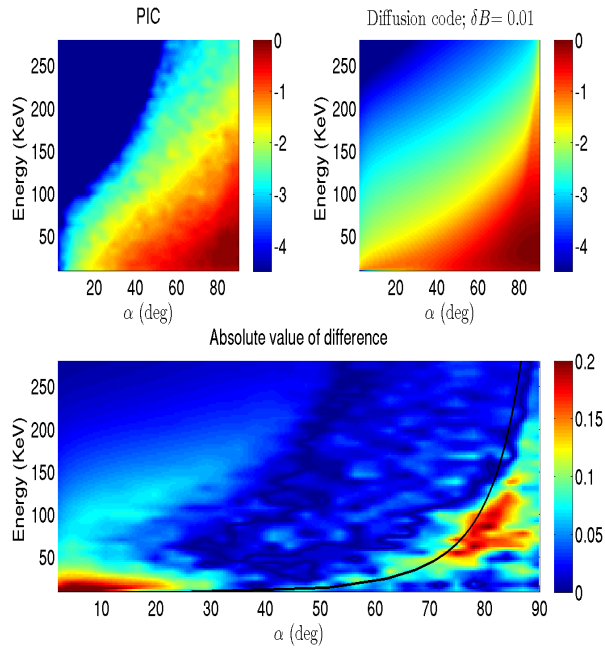}%
 \caption{Comparison between PIC and diffusion code results. Probability density
 functions for PIC (top-left), and diffusion code for the case $\delta B/B_0=1\%$(top-right), at time $T\Omega_e = 1000$.
 The bottom plot shows the absolute value of the difference between the two solutions. Below the black line 
 the resonant condition Eq. (\ref{resonance}) requires $\omega/\Omega_e > 0.5$, and therefore only non-resonant interactions are possible.
 \label{fig:diff_PIC}}
 \end{figure}
It shows the absolute value of the difference between the diffusion code and the PIC results (we remind that the distribution function values range between 0 and 1).
The superposed black line denotes the minimal resonant energy for given pitch angle, calculated via Eq. (\ref{resonance2}),
by using a maximum frequency $\omega_M/\Omega_e=0.5$. 
Very interestingly, for large pitch angles the larger mismatches lie below the black curve, where non-resonant scattering takes place and, as such, 
the diffusion code performs poorly. The small pitch angle region is also quantitatively different,
with the larger mismatch for small energies. This indicates
that the actual particle lifetime might be larger than the one estimated through the quasi-linear diffusion coefficients.

\section*{Discussion}
We have presented a direct comparison between first-principle PIC and
quasi-linear diffusive simulations. The focus has been on one-dimensional PIC simulations of wave-particle interactions between suprathermal electron
and lower-band chorus waves.
In PIC, the waves are self-consistently generated by an initial small population of anisotropic energetic electrons.
This approach does not require any of the assumptions used by quasi-linear theory or test-particle simulations.
The particle diagnostic has been performed on samples of 160,000 PIC particles, resulting in an excellent statistics.
It is important to remind that we have chosen a given value of $\delta B/B_0$, for the calculation of diffusion coefficients,
and that the value that results in the best agreement between the two codes is reached at an early time during the wave growth,
i.e. for $T\Omega_e = 200$.
Although the two approaches give qualitatively similar results, we have highlighted some important differences.
First, the quasi-linear code generally overestimates diffusion for small pitch angles. This
is an important result that implies a reconsideration of the standard estimates of particle lifetime,
which is usually based on the characteristic diffusion times at loss cone angles.
Second, we have presented evidence of non-resonant wave-particle interactions at large pitch angles that, by construction, 
cannot adequately be described in the standard quasi-linear framework. 
In this respect, it would be interesting to test higher-order nonlinear theories, such as the ones described in Ref. \citep{qin09}.\\
In conclusions, our PIC simulations corroborate the long standing viewpoint that diffusion codes are a powerful reduced model for
the study of wave-particle interaction phenomena in the radiation belts, but at the same time,
in view of more quantitative predictions, they solicit an effort to include non-resonant interactions.

%
%


%
%
%
%
%
%
%

\section*{Acknowledgments}
Data used in this paper is available upon request to the author.
\end{document}